\documentclass[%
 reprint,
 amsmath,amssymb,
 aps,
]{revtex4-2}
\usepackage[colorlinks=true,
            linkcolor=blue,
            citecolor=blue,
            urlcolor=blue]{hyperref}
\usepackage{soul}
\usepackage{graphicx}
\usepackage{dcolumn}
\usepackage{bm}

\begin{document}

\title{Halo Structure and Lensing Signatures of a Polytropic Dark Matter Fluid}

\author{Marriam Naeem}
\email{mariyum@mariyumresearch.com}

\affiliation{%
The Women University,\\
Multan, Pakistan
}

\date{\today}

\begin{abstract}
We investigate whether a minimal effective pressure in the dark matter sector can modify nonlinear halo structure while preserving the successful large-scale predictions of the $\Lambda$ cold dark matter ($\Lambda$CDM) model. To explore this possibility, we consider a barotropic relation $P = K \rho^{\gamma}$ with $\gamma = 3/2$, interpreted as an effective coarse-grained closure of the Jeans hierarchy in virialized regions. In this framework, the dark matter fluid remains effectively pressureless at cosmological densities while developing a finite effective sound speed inside collapsed halos.

For $\gamma = 3/2$, the equilibrium halo configurations correspond to the $n = 2$ Lane--Emden solution, which produces finite-radius density profiles with quadratic central flattening. When embedded within the empirical concentration--mass relation of $\Lambda$CDM halos, the resulting core scale exhibits only weak mass dependence across dwarf-to-galaxy mass ranges. For parameter values that generate kiloparsec-scale cores, the background expansion history and the linear growth of density perturbations remain observationally indistinguishable from $\Lambda$CDM, while the present-day Jeans scale remains confined to sub-megaparsec lengths.

We compute the projected surface density and the associated weak-lensing convergence profiles for the polytropic halo model. Relative to mass-matched Navarro--Frenk-White halos, the model predicts a moderate suppression of the central lensing amplitude, while the convergence power spectrum is modified only at sufficiently high multipoles. Therefore, current weak-lensing measurements do not impose significant constraints within the dynamically stable regime $\gamma > 4/3$.

The model introduces a single additional parameter controlling nonlinear pressure support and continuously reduces to collision-free cold dark matter in the limit $K \rightarrow 0$. Therefore, it provides a simple phenomenological framework in which the inner halo structure can be modified without altering the established large-scale behavior of $\Lambda$CDM.

\end{abstract}

\maketitle
\section{Introduction}

The $\Lambda$ cold dark matter ($\Lambda$CDM) paradigm provides a remarkably successful description of the large-scale Universe. Observations of the cosmic microwave background, baryon acoustic oscillations, galaxy clustering, and weak gravitational lensing are all consistent with a cosmological model in which structure formation proceeds through gravitational instability in a pressureless dark matter component supplemented by a cosmological constant \cite{Planck2018,WeinbergReview}. On scales larger than individual galaxies, collisionless cold dark matter accurately reproduces the observed matter power spectrum and halo mass function.

However, on galactic and sub-galactic scales, the internal structure of dark matter halos remains an open problem. Numerical simulations of collisionless cold dark matter generically predict cuspy inner density profiles, commonly described by the Navarro--Frenk--White (NFW) profile \cite{NavarroFrenkWhite}. In contrast, observational studies of dwarf and low-surface-brightness galaxies often favor shallower, cored central density distributions \cite{deBlokReview}. Although baryonic feedback processes may partially alleviate this discrepancy, it remains unclear whether such mechanisms can fully account for the diversity of inferred core structures \cite{BullockBoylanKolchin}.These considerations motivate the exploration of controlled extensions of the standard pressureless dark matter assumption that preserve the successful large-scale predictions of $\Lambda$CDM.

In the standard cosmological treatment, dark matter is modeled as a pressureless perfect fluid across all astrophysical regimes. While this approximation is well justified in the linear regime and for the background expansion history, it represents an idealization of a more general possibility in which dark matter may develop a small but finite effective pressure in sufficiently dense environments. A minimal phenomenological realization of this idea is provided by a barotropic equation of state
\begin{equation}
P = K \rho^{\gamma},
\end{equation}
which reduces to pressureless behavior at cosmological densities while introducing nonlinear pressure support within collapsed halos.

In this work, we interpret this relation as an effective coarse-grained closure of the collisionless Jeans hierarchy in virialized halo interiors. This approach provides a minimal extension of collisionless cold dark matter that does not rely on particle self-interactions, quantum coherence, or modifications of gravitational dynamics. Instead, it isolates the role of a density-dependent effective pressure in shaping the halo structure while preserving the successful large-scale predictions of the $\Lambda$CDM model.

The central objective of this work is to test whether a minimal effective pressure arising from a coarse-grained Jeans closure can generate cored halo structures while preserving the background expansion and linear perturbation predictions of the $\Lambda$CDM model.

To address this question, we consider a polytropic dark matter fluid with $\gamma = 3/2$, corresponding to a polytropic index $n = 2$, which satisfies the stability condition $\gamma > 4/3$ for self-gravitating systems. We analyze equilibrium halo configurations using the Lane--Emden formalism, derive the scaling relations for the core structure, and examine the implications for linear perturbation growth and the Jeans scale. We further compute projected surface density and weak-lensing convergence profiles to assess potential observational signatures. Throughout this work, we restrict attention to parameter ranges that remain consistent with current cosmological constraints.

\subsection{Relation to Previous Polytropic and Dark Sector Models}

Polytropic descriptions of self-gravitating matter have a long history in stellar structure theory and have occasionally been applied to dark matter halos as phenomenological models \cite{SaxtonFerreras,Novotny,CalvoPolytrope,ChavanisPolytrope}. In many such studies, the polytropic equation of state is introduced either as a fitting function for equilibrium configurations or as an effective description motivated by specific microphysical scenarios. The framework considered here differs in that the polytropic relation is interpreted as a minimal thermodynamic extension of collisionless cold dark matter and is treated consistently across background cosmology, linear perturbation theory, halo equilibrium, and projected lensing observables. This unified treatment allows a direct connection between halo structure and cosmological dynamics within a single effective description.

Ultralight scalar or Bose–Einstein condensate (BEC) dark matter models also generate cored halo structures \cite{FDM_Hu,FDM_Review}. In these scenarios, pressure support originates from quantum mechanical gradient energy, producing solitonic cores and characteristic interference patterns. The core size is determined by the particle mass and follows scaling relations distinct from classical polytropic solutions. In contrast, the present model is entirely classical at macroscopic scales: the pressure arises from a barotropic relation $P = K\rho^\gamma$, and equilibrium configurations are governed by the Lane–Emden equation without invoking quantum coherence or particle mass tuning. As a result, the model does not predict wave interference effects or density granularity.

Self-interacting dark matter (SIDM) provides another mechanism for core formation through collisional heat transport in halo centers \cite{SIDM_SpergelSteinhardt,SIDM_Tulin}. In such models, particle scattering leads to thermalization and gravothermal evolution, with core properties determined by the interaction cross section. While SIDM can successfully produce cored profiles, it introduces additional particle physics parameters and requires assumptions about the nature and strength of dark matter interactions. By contrast, the present framework does not rely on microscopic interactions. Core formation arises directly from hydrostatic balance with a density-dependent effective pressure, thereby avoiding cross-section tuning.

Recent developments have also explored more general compact configurations, including anisotropic stellar models, geometrically deformed compact objects, and extensions involving nontrivial spacetime structure \cite{Anisotropic_Instability_2025,BTZ_Deformed_2026,DarkEnergy_Star_2026,Nonmetricity_Stars_2025,Torsion_Stars_2025,Axial_Config_2024}.

These approaches often modify the stress–energy tensor or the gravitational sector to obtain equilibrium solutions with distinct structural properties. Similarly, exotic compact objects such as wormholes supported by dark matter or modified energy conditions have been investigated as alternative gravitational configurations \cite{FDM_Wormhole_2026,DM_Wormhole_2026,Einasto_Dymnikova_2025}.

While these models provide valuable theoretical insights, they typically extend beyond minimal modifications of the cold dark matter paradigm. In contrast, the present work retains standard gravitational dynamics and modifies only the effective thermodynamic description of the dark matter component.

Additional studies have also examined the interplay between dark matter and stellar interiors in extended gravitational frameworks \cite{DM_StellarInterior_2026}.

Effective fluid descriptions of the dark sector have also been considered in the context of viscous cosmology \cite{Viscous_Cosmology_2025}, where dissipative stresses contribute to the effective pressure and can influence cosmic expansion and structure formation. In such models, the pressure generally depends on velocity gradients and introduces entropy production through irreversible processes. The approach adopted here differs fundamentally in that it is non-dissipative. The pressure is specified by a barotropic relation$P = K\rho^\gamma$ and represents an effective, non-dissipative closure of the collisionless Jeans hierarchy. This distinction enables a direct and consistent link between halo equilibrium structure, the effective sound speed, and linear perturbation dynamics.

Overall, the framework considered in this work occupies a distinct position among dark sector models. It introduces a single phenomenological parameter that governs a density-dependent effective pressure, without invoking new particle species, modified gravity, or dissipative processes.The model therefore provides a minimal and controlled extension of collisionless cold dark matter, allowing us to isolate the role of nonlinear pressure support in shaping halo interiors while preserving the well-established large-scale predictions of the $\Lambda$CDM paradigm \cite{BullockBoylanKolchin}.

\section{Dark-Sector Thermodynamic Framework}

\subsection{Effective Barotropic Description}

We consider a dark matter component described at macroscopic scales as a self-gravitating perfect fluid with stress–energy tensor
\begin{equation}
T_{\mu\nu} = (\rho + P)u_\mu u_\nu + P g_{\mu\nu},
\end{equation}
where $\rho$ is the energy density, $P$ is the pressure, and $u^\mu$ is the four-velocity satisfying $u^\mu u_\mu = -1$. We assume a barotropic equation of state of the form
\begin{equation}
P = K \rho^{\gamma},
\label{eq:polytrope}
\end{equation}
with constant parameters $K > 0$ and $\gamma > 1$. Equation~(\ref{eq:polytrope}) reduces to the standard cold dark matter limit when $K \rightarrow 0$, in which case $P \rightarrow 0$ at all densities.

The adiabatic sound speed follows directly from
\begin{equation}
c_s^2 \equiv \frac{dP}{d\rho} = K \gamma \rho^{\gamma - 1}.
\label{eq:soundspeed}
\end{equation}
For sufficiently small values of $K$ and for cosmological background densities $\bar{\rho}$, the sound speed satisfies $c_s^2 \ll 1$. In this limit, the background expansion and the growth of linear density perturbations closely reproduce the predictions of $\Lambda$CDM. The pressure contribution becomes dynamically relevant only in regions where the density approaches halo-scale values.
\paragraph{Relation to viscous effective fluid descriptions.}
Effective fluid descriptions of the dark sector have also been explored in the context of viscous cosmology, in which the stress–energy tensor includes dissipative contributions. In such models, the effective pressure typically takes the form
\begin{equation}
P_{\mathrm{eff}} = P - \zeta \, \nabla_\mu u^\mu
\end{equation}
where $\zeta$ denotes the bulk viscosity coefficient and $\nabla_\mu u^\mu$ is the covariant velocity divergence. This contribution introduces entropy production and characterizes irreversible, non-equilibrium dynamics, which can affect both the background expansion and the growth of structure.

The framework considered in this work differs fundamentally from viscous models.The pressure is specified by a barotropic relation $P = K \rho^{\gamma}$ and represents an effective, non-dissipative closure of the collisionless Jeans hierarchy. No velocity-gradient terms are introduced, and the dynamics remain conservative at the macroscopic level. As a result, the same equation of state consistently governs both equilibrium halo structure and linear perturbation evolution. This distinction allows a direct connection between the effective sound speed, the Jeans scale, and the internal structure of dark matter halos, without invoking dissipative processes or entropy generation.

\subsection{Relation to the Polytropic Index}

It is convenient to rewrite Eq.~(\ref{eq:polytrope}) in the standard polytropic form
\begin{equation}
P = K \rho^{1 + 1/n},
\end{equation}
where the polytropic index $n$ is related to $\gamma$ by
\begin{equation}
\gamma = 1 + \frac{1}{n},
\qquad
n = \frac{1}{\gamma - 1}.
\label{eq:gamma_n_relation}
\end{equation}

In the following, we focus on the case
\begin{equation}
\gamma = 1.5,
\qquad
n = 2.
\end{equation}
This choice satisfies several physical considerations:

(i) \emph{Dynamical stability;} For self-gravitating configurations, stability against radial perturbations requires $\gamma > 4/3$. The choice $\gamma = 1.5$ satisfies this condition and therefore allows stable equilibrium solutions.

(ii) \emph{Finite-radius halos;} For polytropes with $n < 5$, solutions of the Lane–Emden equation possess a finite radius. Since $n = 2 < 5$, the corresponding halos terminate at a finite outer boundary without requiring an imposed truncation.

(iii) \emph{Mass-dependent core structure.} Unlike the special case $\gamma = 2$ ($n = 1$), which produces a core radius independent of the central density, the choice $n = 2$ leads to equilibrium scales that explicitly depend on $\rho_c$. This allows the core radius to vary with halo mass.

(iv) \emph{Controlled suppression of small-scale structure.} From Eq.~(\ref{eq:soundspeed}), the sound speed scales as $c_s^2 \propto \rho^{\gamma - 1} = \rho^{1/2}$. For cosmological densities $\bar{\rho}$ this contribution remains negligible, while in high-density halo interiors it becomes dynamically relevant without producing excessive Jeans-scale suppression in the linear regime.

\subsection{Hydrostatic Equilibrium and Dimensionless Formulation}

Under spherical symmetry, hydrostatic equilibrium requires
\begin{equation}
\frac{dP}{dr} = - \rho(r) \frac{G M(r)}{r^2},
\label{eq:hydro}
\end{equation}
where the enclosed mass satisfies
\begin{equation}
\frac{dM}{dr} = 4 \pi r^2 \rho(r).
\label{eq:masscontinuity}
\end{equation}

Substituting Eq.~(\ref{eq:polytrope}) into Eq.~(\ref{eq:hydro}) yields
\begin{equation}
K \gamma \rho^{\gamma - 1} \frac{d\rho}{dr}
=
- \rho \frac{G M}{r^2}.
\end{equation}

Introducing dimensionless variables
\begin{equation}
\rho(r) = \rho_c \theta^n,
\qquad
r = a \xi,
\end{equation}
with the scale parameter
\begin{equation}
a^2 = \frac{(n+1)K}{4\pi G} \rho_c^{1/n - 1},
\label{eq:scaleradius}
\end{equation}
one obtains the Lane–Emden equation
\begin{equation}
\frac{1}{\xi^2}\frac{d}{d\xi}
\left(
\xi^2 \frac{d\theta}{d\xi}
\right)
=
- \theta^n.
\label{eq:laneemden}
\end{equation}

For $n = 2$, Eq.~(\ref{eq:laneemden}) admits solutions with finite radius $\xi_1 \approx 4.35$, which defines the halo boundary
\begin{equation}
R_{\rm halo} = a \, \xi_1.
\end{equation}

The scale parameter $a$ depends on the central density as
\begin{equation}
a \propto \rho_c^{(1/n - 1)/2}
=
\rho_c^{-1/4},
\end{equation}
indicating that characteristic core scales vary with halo density. This contrasts with the $n = 1$ case, where $a$ is independent of $\rho_c$.

\subsection{Cosmological Limit}

In a homogeneous background with density $\bar{\rho}(t)$, the equation-of-state parameter is
\begin{equation}
w \equiv \frac{P}{\rho}
=
K \bar{\rho}^{\gamma - 1}.
\end{equation}

For cosmological densities $\bar{\rho} \sim 10^{-29}\,{\rm g\,cm^{-3}}$ and parameter values calibrated to produce kiloparsec-scale halo cores, one finds $w \ll 1$. Therefore, the background expansion history remains indistinguishable from $\Lambda$CDM at the current observational precision. The pressure contribution becomes dynamically significant only in nonlinear overdensities where $\rho \gg \bar{\rho}$.

Equation~(\ref{eq:polytrope}) should therefore be interpreted as an effective coarse-grained closure of the collisionless Jeans hierarchy in virialized halos.

\subsection*{Physical Interpretation of the Effective Polytropic Closure}

The barotropic relation $P = K\rho^\gamma$ is introduced here as an effective macroscopic closure rather than as a fundamental microphysical equation of state. In collisionless cold dark matter, the phase-space distribution function $f(\mathbf{x},\mathbf{v},t)$ obeys the collisionless Boltzmann (Vlasov) equation. Taking velocity moments of this equation generates a hierarchy involving higher-order moments of the velocity distribution, which does not close without additional assumptions.

In approximately virialized halo interiors, numerical simulations indicate that the velocity dispersion tensor becomes nearly isotropic and varies smoothly toward the halo center. Under these conditions, the coarse-grained stress tensor can be approximated as
\begin{equation}
P \simeq \rho \sigma^2 ,
\label{eq:P_sigma_relation}
\end{equation}
where $\sigma^2$ denotes the one-dimensional velocity dispersion.

For a self-gravitating equilibrium core with characteristic mass $M$ and size $R$, the virial relation implies
\begin{equation}
\sigma^2 \sim \frac{GM}{R}.
\label{eq:virial_scaling}
\end{equation}
Using $M \sim \rho R^3$ for a cored configuration gives
\begin{equation}
\sigma^2 \sim G\rho R^2.
\label{eq:sigma_density_relation}
\end{equation}

For the $n=2$ Lane--Emden solution derived below, the characteristic core scale varies with density approximately as
\begin{equation}
R \propto \rho^{-1/4}.
\label{eq:R_density_scaling}
\end{equation}
Substituting Eq.~(\ref{eq:R_density_scaling}) into Eq.~(\ref{eq:sigma_density_relation}) yields
\begin{equation}
\sigma^2 \propto \rho^{1/2}.
\end{equation}
Combining this relation with Eq.~(\ref{eq:P_sigma_relation}) leads to
\begin{equation}
P \propto \rho^{3/2}.
\end{equation}

The polytropic index
\begin{equation}
\gamma = \frac{3}{2}
\end{equation}
can therefore be interpreted as a natural density-dependent closure of the coarse-grained Jeans hierarchy in quasi-equilibrium halo interiors. Importantly, this argument does not rely on microscopic thermodynamic equilibrium or particle self-interactions. Instead, it represents an effective description of collisionless dynamics in the nonlinear regime.

To provide an observational consistency check for this interpretation, we reconstructed the effective pressure--density relation using galaxy rotation curves from the SPARC database \cite{SPARC}. In this reconstruction, the effective pressure is defined as $P = \rho \sigma^2$, where $\sigma$ denotes the velocity dispersion inferred from the halo dynamics. The reconstructed relation follows approximately a power law $P \propto \rho^{0.9}$, indicating that galaxy halos behave close to the isothermal limit ($\gamma \approx 1$).This empirical result suggests that the polytropic relation should be interpreted as an effective phenomenological closure for halo equilibrium rather than a fundamental microphysical equation of state.

\section{Halo Equilibrium}

\subsection{Hydrostatic Equilibrium}

We consider a spherically symmetric, self-gravitating configuration described by the barotropic equation of state
\begin{equation}
P = K \rho^{\gamma},
\qquad
\gamma = 1.5,
\end{equation}
corresponding to a polytropic index $n = 2$. In the non-relativistic regime appropriate for galactic halos, equilibrium is governed by the Newtonian hydrostatic condition
\begin{equation}
\frac{dP}{dr}
=
- \rho(r) \frac{G M(r)}{r^2},
\label{eq:hydrostatic}
\end{equation}
together with mass continuity,
\begin{equation}
\frac{dM}{dr}
=
4\pi r^2 \rho(r).
\label{eq:masscontinuityIII}
\end{equation}

Substituting $P = K \rho^{1+1/n}$ into Eq.~(\ref{eq:hydrostatic}) and introducing the dimensionless variables
\begin{equation}
\rho(r) = \rho_c \theta^n,
\qquad
r = a \xi,
\end{equation}
with
\begin{equation}
a^2 = \frac{(n+1)K}{4\pi G} \rho_c^{1/n - 1},
\label{eq:a_def}
\end{equation}
yields the Lane–Emden equation\cite{LaneEmdenOriginal,Chandrasekhar}
\begin{equation}
\frac{1}{\xi^2}
\frac{d}{d\xi}
\left(
\xi^2 \frac{d\theta}{d\xi}
\right)
=
- \theta^n,
\qquad n=2.
\label{eq:LE_n2}
\end{equation}

The solution satisfying regular central boundary conditions,
\begin{equation}
\theta(0)=1,
\qquad
\theta'(0)=0,
\end{equation}
vanishes at a finite radius $\xi_1 \approx 4.35$, which defines the halo boundary
\begin{equation}
R_{\rm halo} = a \xi_1.
\end{equation}

Since $n = 2 < 5$, the solution possesses finite extent without requiring an externally imposed truncation. The total halo mass is therefore
\begin{equation}
M_{\rm halo}
=
4\pi a^3 \rho_c
\left[
- \xi^2 \theta'(\xi)
\right]_{\xi=\xi_1}.
\label{eq:mass_total}
\end{equation}

\subsection{Core Structure}

Near the origin, the Lane–Emden solution admits the expansion
\begin{equation}
\theta(\xi)
=
1 - \frac{\xi^2}{6}
+ \mathcal{O}(\xi^4),
\end{equation}
which implies
\begin{equation}
\rho(r)
=
\rho_c
\left(
1 - \frac{r^2}{6a^2}
\right)^2
+
\mathcal{O}(r^4).
\end{equation}

The density therefore remains finite at $r = 0$ and exhibits quadratic central flattening. We define the characteristic core radius as $R_c \sim a$.

From Eq.~(\ref{eq:a_def}) with $n = 2$,
\begin{equation}
a \propto K^{1/2} \rho_c^{-1/4}.
\label{eq:a_scaling}
\end{equation}

\subsection{Comparison with NFW Halo Profiles}

To illustrate the structural differences relative to the standard $\Lambda$CDM halo model, we compare the density profile predicted by the $n=2$ polytropic solution with the widely used Navarro–Frenk–White (NFW) profile \cite{NavarroFrenkWhite}. 

The NFW density profile takes the form
\begin{equation}
\rho_{\rm NFW}(r) =
\frac{\rho_s}{(r/r_s)(1+r/r_s)^2},
\end{equation}
where $\rho_s$ and $r_s$ denote the characteristic density and scale radius. This profile exhibits a central cusp with inner slope
\begin{equation}
\rho(r) \propto r^{-1}
\quad (r \rightarrow 0).
\end{equation}

In contrast, the polytropic halo solution derived from the Lane–Emden equation approaches a constant central density. As shown in Eq.~(\ref{eq:a_scaling}), the density near the origin behaves as
\begin{equation}
\rho(r) \approx \rho_c
\left(1-\frac{r^2}{6a^2}\right)^2,
\end{equation}
implying
\begin{equation}
\lim_{r\to0} \rho(r) = \rho_c .
\end{equation}

the polytropic model therefore produces a finite-density core rather than a central cusp. At larger radii the density profile gradually steepens, approaching behavior similar to the outer regions of NFW halos where the gravitational potential is dominated by the total halo mass.

\begin{figure}
    \centering
    \includegraphics[width=0.5\linewidth]{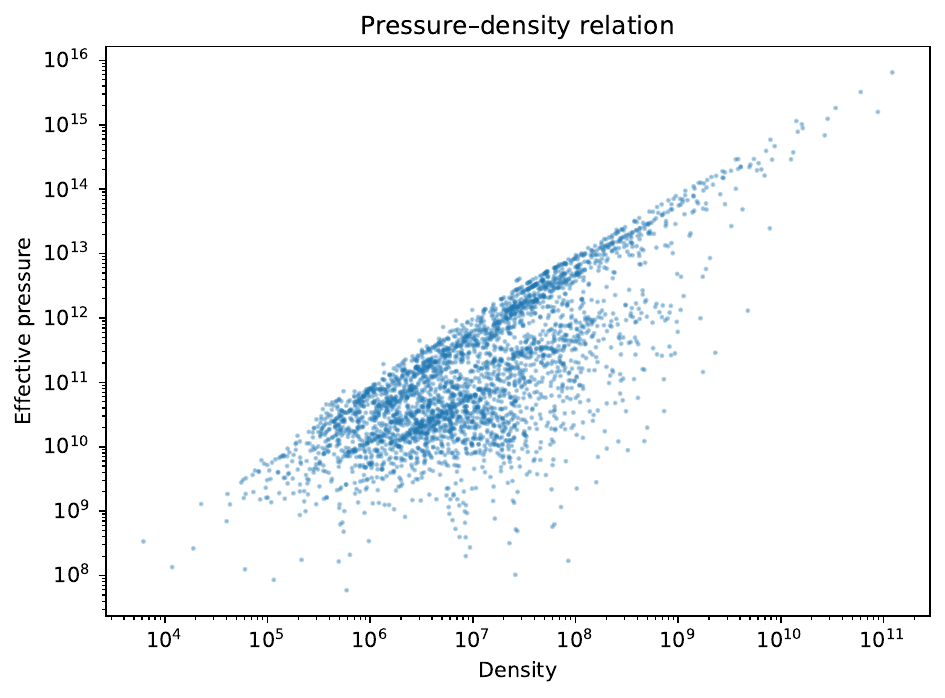}
    \caption{
Comparison between the density profiles of the $n=2$ polytropic halo model and the Navarro–Frenk–White (NFW) profile for halos of identical virial mass. 
The polytropic solution exhibits a finite-density core, while the NFW profile develops a central cusp with $\rho \propto r^{-1}$. 
At larger radii the profiles approach similar asymptotic behavior.
}
    \label{fig:nfw_comparison}
\end{figure}

Figure~\ref{fig:nfw_comparison} illustrates the resulting density profiles for halos with identical virial mass. The comparison highlights the main structural difference between the two models: the NFW profile develops a cuspy inner slope, whereas the polytropic model generates a smooth, finite-density core.

\subsection{Isolated Polytropic Scaling}

For an isolated Lane–Emden configuration, Eq.~(\ref{eq:mass_total}) implies
\begin{equation}
M_{\rm halo} \propto a^3 \rho_c.
\end{equation}

Using Eq.~(\ref{eq:a_scaling}),
\begin{equation}
M_{\rm halo}
\propto
K^{3/2} \rho_c^{1/4}.
\end{equation}

Eliminating $\rho_c$ gives
\begin{equation}
R_c \sim a \propto K^{3/5} M_{\rm halo}^{-1/5}.
\label{eq:Rc_isolated}
\end{equation}

Equation~(\ref{eq:Rc_isolated}) therefore describes the scaling relation for isolated equilibrium polytropic halos and serves primarily as a structural reference for the more realistic cosmological halos considered below.

\subsection{Embedding in $\Lambda$CDM Halos}

Cosmological halos are not isolated polytropes; their internal densities correlate with the empirical concentration–mass relation
\begin{equation}
c(M) \propto M^{-\beta},
\end{equation}
with $\beta \simeq 0.08$--$0.12$ at $z=0$.

Since the characteristic central density scales approximately as $\rho_c \propto c^3$, one obtains
\begin{equation}
\rho_c(M) \propto M^{-3\beta}.
\end{equation}

Using $R_c \propto \rho_c^{-1/4}$ then yields
\begin{equation}
R_c(M)
\propto
M^{3\beta/4}.
\label{eq:Rc_cosmo}
\end{equation}

For $\beta \simeq 0.1$, this gives
\begin{equation}
R_c(M) \propto M^{0.075},
\end{equation}
corresponding to a very weak mass dependence across dwarf-to-galaxy scales.

\subsection{Core Surface Density}

An observationally relevant quantity in halo phenomenology is the central surface density,
\begin{equation}
\Sigma_0 \equiv \rho_c R_c ,
\end{equation}
where $\rho_c$ denotes the characteristic central density and $R_c$ the associated core scale.

Embedding the equilibrium solution within the empirical concentration--mass relation
$c(M) \propto M^{-\beta}$ implies
\begin{equation}
\rho_c(M) \propto M^{-3\beta}.
\end{equation}
Since the core scale satisfies
\begin{equation}
R_c \propto \rho_c^{-1/4},
\end{equation}
it follows that
\begin{equation}
R_c(M) \propto M^{3\beta/4}.
\end{equation}

The central surface density therefore scales as
\begin{equation}
\Sigma_0(M)
= \rho_c R_c
\propto
M^{-3\beta} \, M^{3\beta/4}
=
M^{-9\beta/4}.
\label{eq:Sigma_scaling}
\end{equation}

For representative values $\beta \simeq 0.1$, consistent with numerical
simulations of $\Lambda$CDM halos, this reduces to
\begin{equation}
\Sigma_0(M)
\propto
M^{-0.225},
\end{equation}
indicating a weak decline of central surface density with increasing halo mass.

\begin{figure}
    \centering
    \includegraphics[width=0.5\linewidth]{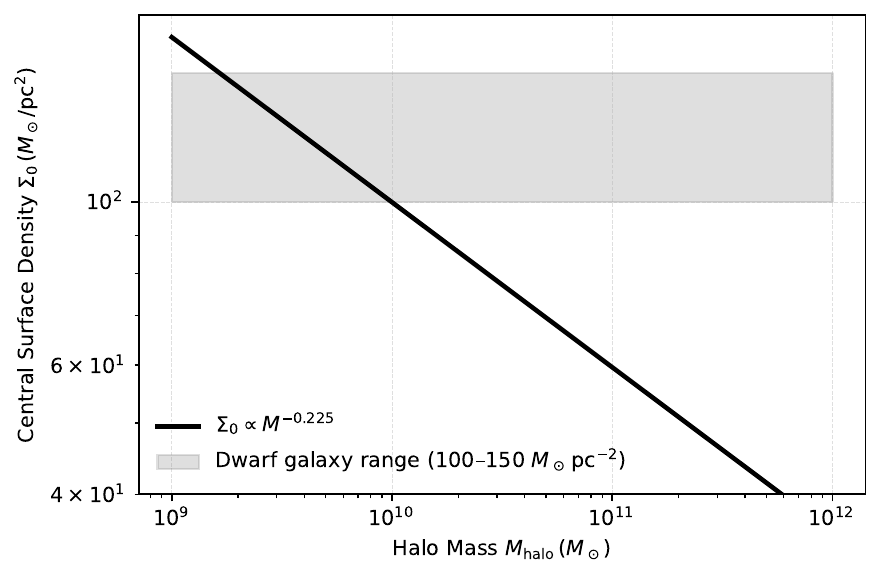}
       \caption{
    Central surface density $\Sigma_0$ as a function of halo mass for the polytropic model (solid curve), normalized at $10^{10} M_\odot$. 
    The shaded region indicates the approximate range $100$--$150\,M_\odot\,\mathrm{pc}^{-2}$ reported for dwarf galaxies in the literature. 
    The predicted mass dependence is weak and varies smoothly across dwarf-to-galaxy mass scales.
    }  
    \label{fig:sigma_mass}
\end{figure}

To illustrate the magnitude of this dependence, we adopt a normalization at
$M_0 = 10^{10} M_\odot$ such that
$\Sigma_0(M_0) = 100\,M_\odot\,\mathrm{pc}^{-2}$,
consistent with typical dwarf-scale estimates reported in the literature.
The resulting mass dependence is shown in Fig.~\ref{fig:sigma_mass}.

Across several decades in halo mass, the predicted variation remains modest and
comparable to the observational scatter of empirical central surface density
estimates. The model therefore does not introduce pronounced mass-dependent
departures from phenomenologically inferred core properties, while allowing
moderate deviations from strict universality.

\subsection{Comparison with Empirical Central Surface Density Measurements}

An observationally discussed quantity in the context of halo core phenomenology is the central surface density,
\begin{equation}
\Sigma_0 \equiv \rho_c R_c .
\end{equation}
Donato \textit{et al.}~\cite{Donato2009} reported that a wide range of galaxies, from dwarf spheroidals to spirals, exhibit central surface densities clustering around 
$\Sigma_0 \sim 100\,M_\odot\,\mathrm{pc}^{-2}$, with relatively modest scatter. 
This apparent near-universality has been interpreted as a potential clue to dark matter microphysics.

Subsequent analyses have questioned the strict constancy of this relation and suggested the presence of weak mass dependence once systematic effects, profile choices, and sample selection are taken into account 
\cite{Burkert2015,DiPaolo2019}. 
Current data therefore permit either approximate universality or a shallow scaling within observational uncertainties.

In the present framework, embedding the $n=2$ polytropic solution within the empirical concentration--mass relation yields
\begin{equation}
\Sigma_0(M) \propto M^{-9\beta/4},
\end{equation}
which for representative $\beta \simeq 0.1$ gives
\begin{equation}
\Sigma_0(M) \propto M^{-0.225}.
\end{equation}
Across dwarf-to-galaxy mass scales spanning several decades, this corresponds to variations of at most a factor of a few. 
When normalized at $M_0 = 10^{10} M_\odot$ to $\Sigma_0(M_0)=100\,M_\odot\,\mathrm{pc}^{-2}$, the predicted values remain within the observational scatter reported in the current datasets.

The model therefore does not predict the strict universality of the central surface density, but yields only a mild mass dependence consistent with the present empirical uncertainties. 
Future high-precision determinations of halo core parameters across a broad mass range may help discriminate between exact constancy and weak scaling of $\Sigma_0(M)$.

\subsection{Inner Logarithmic Slope}

For small $r$,
\begin{equation}
\frac{d \ln \rho}{d \ln r}
=
- \frac{2r^2}{3a^2}
+
\mathcal{O}(r^4),
\end{equation}
so that
\begin{equation}
\lim_{r \to 0}
\frac{d \ln \rho}{d \ln r}
=
0.
\end{equation}

The transition from a flat core to a declining profile occurs smoothly over $r \sim a$.

\subsection{Sensitivity to $\gamma$}

For general $\gamma$, the Lane–Emden scale parameter satisfies
\begin{equation}
a^2 \propto K \rho_c^{\gamma-2}.
\end{equation}

Embedding within $\rho_c(M)\propto M^{-3\beta}$ gives
\begin{equation}
R_c(M) \propto M^{\frac{3\beta}{2}(2-\gamma)},
\end{equation}
and therefore
\begin{equation}
\Sigma_0(M) \propto M^{-\frac{3\beta\gamma}{2}}.
\end{equation}

For $\gamma$ in the range $1.4 \lesssim \gamma \lesssim 1.6$ and $\beta \simeq 0.1$, the exponent remains small, indicating that the qualitative halo structure is not highly sensitive to modest variations in $\gamma$.

We note that the dynamical stability of self-gravitating polytropes requires $\gamma > 4/3$, and all parameter values considered here satisfy this condition.

The $n=2$ polytropic solution yields:

\begin{itemize}
\item Finite total mass without imposed truncation,
\item Finite central density with quadratic inner flattening,
\item Weak mass dependence when embedded in $\Lambda$CDM halos,
\item Controlled deviations from NFW structure,
\item Continuous recovery of collisionless behavior as $K \rightarrow 0$.
\end{itemize}

These properties arise from hydrostatic balance with a density-dependent effective sound speed and do not require modifications to the gravitational dynamics. These analytic results provide a simple theoretical framework for understanding cored halo structures and motivate the comparison with the observational rotation curves discussed in the following.

\section{Cosmological Consistency}

\subsection{Background Evolution}

We first verify that the polytropic extension does not modify the homogeneous expansion history at observable precision. In a spatially flat FLRW spacetime, energy conservation implies
\begin{equation}
\dot{\rho} + 3H(\rho + P) = 0,
\end{equation}
where $H = \dot{a}/a$ is the Hubble parameter. Using the equation of state
\begin{equation}
P = K \rho^{\gamma},
\qquad
\gamma = 1.5,
\end{equation}
the effective equation-of-state parameter becomes
\begin{equation}
w \equiv \frac{P}{\rho}
=
K \rho^{\gamma - 1}
=
K \rho^{1/2}.
\label{eq:w_background}
\end{equation}

For cosmological background densities 
$\bar{\rho}_0 \simeq 3H_0^2 \Omega_m / 8\pi G
\sim 10^{-29}\,\mathrm{g\,cm^{-3}}$
and for parameter values $K$ that generate kiloparsec-scale cores in dwarf halos, one obtains
\begin{equation}
w_0 \lesssim 10^{-6}.
\end{equation}

Thus $w \ll 1$ at all redshifts of interest, and the background density evolves as
\begin{equation}
\bar{\rho}(a) \simeq \rho_0 a^{-3}
\end{equation}
to excellent approximation. The resulting expansion history is therefore observationally indistinguishable from $\Lambda$CDM within current precision.

\subsection{Linear Density Perturbations}

In the Newtonian regime appropriate for subhorizon modes, linear density perturbations satisfy\cite{MaBertschinger}
\begin{equation}
\ddot{\delta}
+
2H\dot{\delta}
=
4\pi G \bar{\rho} \, \delta
-
\frac{c_s^2 k^2}{a^2} \delta,
\label{eq:delta_evolution}
\end{equation}
where the adiabatic sound speed is
\begin{equation}
c_s^2
=
\frac{dP}{d\rho}
=
K \gamma \rho^{\gamma - 1}
=
1.5 K \rho^{1/2}.
\label{eq:cs2_linear}
\end{equation}

The corresponding comoving Jeans wavenumber is
\begin{equation}
k_J^2
=
\frac{4\pi G \bar{\rho} a^2}{c_s^2}.
\label{eq:jeans}
\end{equation}

Using Eq.~(\ref{eq:cs2_linear}) gives
\begin{equation}
k_J^2 \propto \bar{\rho}^{1/2},
\qquad
k_J \propto a^{-3/4},
\end{equation}
since $\bar{\rho} \propto a^{-3}$.

For parameter values yielding $R_c \sim 1\,\mathrm{kpc}$ in dwarf halos, the present-day Jeans scale evaluates to
\begin{equation}
\lambda_J \equiv \frac{2\pi}{k_J}
\sim 0.1\text{--}0.2\,\mathrm{Mpc},
\end{equation}
corresponding to a characteristic mass scale
\begin{equation}
M_J \sim 10^8\text{--}10^9\,M_\odot.
\end{equation}

This scale lies below those probed by large-scale galaxy clustering and is comparable to, but does not exceed, the smallest scales currently constrained by Lyman-$\alpha$ forest measurements. Consequently, linear growth for modes satisfying $k \ll k_J$ reduces to
\begin{equation}
\ddot{\delta}
+
2H\dot{\delta}
=
4\pi G \bar{\rho} \, \delta,
\end{equation}
recovering the standard $\Lambda$CDM growth equation on observable cosmological scales.
\subsection{Matter Power Spectrum}

Using the linear perturbation equation (\ref{eq:delta_evolution}),we compute the matter power spectrum for the polytropic dark matter model and compare it with the standard $\Lambda$CDM prediction. The spectra are obtained using the CLASS Boltzmann solver.

For parameter values that produce kiloparsec-scale cores in dwarf halos, the effective sound speed at cosmological densities remains extremely small. As a result, the linear growth of perturbations proceeds in essentially the same manner as in the standard cold dark matter scenario.

Consequently, the resulting matter power spectrum is practically indistinguishable from the $\Lambda$CDM prediction across the range of scales probed by current large-scale structure observations ($k \lesssim 10\,h\,\mathrm{Mpc}^{-1}$). This confirms that the effective pressure introduced by the polytropic closure does not modify the background expansion or the linear growth of the cosmological structure \cite{StructureFormation}.

\begin{figure}
    \centering
    \includegraphics[width=0.5\linewidth]{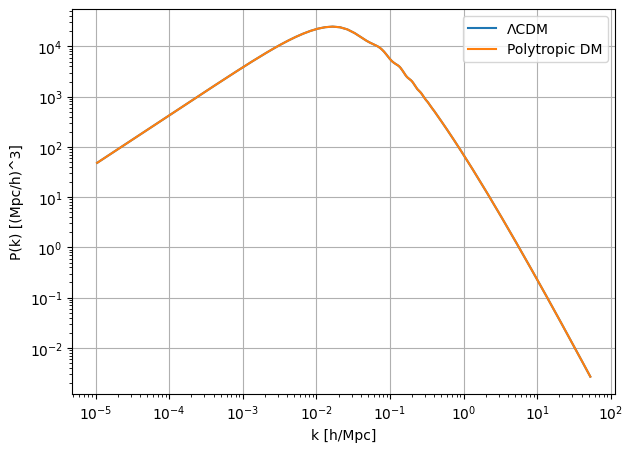}
    \caption{
Matter power spectrum for the polytropic dark matter model compared with the standard $\Lambda$CDM prediction, computed using the CLASS Boltzmann solver. 
The two spectra overlap across the range of cosmological scales probed by current large-scale structure observations, indicating that the effective pressure associated with the polytropic closure does not modify the linear growth of perturbations.
}
\label{fig:powerspectrum}
\end{figure}

The resulting spectra are shown in Fig.~\ref{fig:powerspectrum}, where the prediction of the polytropic model is compared directly with the $\Lambda$CDM case. The two curves overlap across observable cosmological scales, indicating that the proposed framework preserves the successful large-scale predictions of the standard model while allowing modifications to arise only in nonlinear halo interiors.

\subsection{Rotation Curve Slope Distribution}

To test the assumption of approximately constant velocity dispersion in halo interiors, we examine the distribution of logarithmic rotation-curve slopes derived from the SPARC galaxy sample\cite{SPARC}. 

For each galaxy we compute the quantity
\begin{equation}
\alpha \equiv \frac{d\ln v}{d\ln r},
\end{equation}
where $v(r)$ denotes the observed rotation velocity. Values of $\alpha$ close to zero indicate approximately flat rotation curves and therefore nearly constant velocity dispersion in the inner halo region.

\begin{figure}
    \centering
    \includegraphics[width=0.5\linewidth]{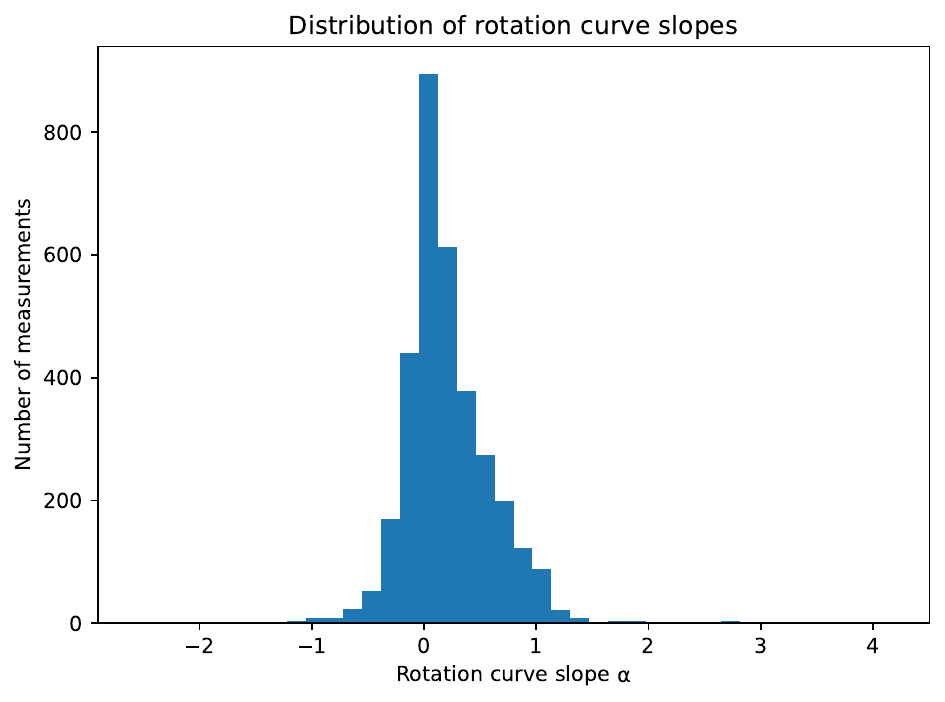}
    \caption{
Distribution of logarithmic rotation-curve slopes $\alpha = d\ln v/d\ln r$ derived from SPARC galaxies. 
The distribution peaks near $\alpha \approx 0$, indicating approximately flat rotation curves and supporting the assumption of slowly varying velocity dispersion in halo interiors.
}    
    \label{fig:slope_distribution}
\end{figure}

The resulting distribution of slopes is shown in Fig.~\ref{fig:slope_distribution} values are strongly clustered around $\alpha \approx 0$, indicating that many galaxy halos are close to isothermal equilibrium in their central regions.

This observational behavior supports the assumption of slowly varying velocity dispersion used in the effective fluid description of halo dynamics adopted in this work.

\subsection{Summary of the Cosmological Regime}

The effective pressure contribution scales as $\rho^{1/2}$ and is therefore negligible in both the homogeneous background and the linear perturbation regime. Its dynamical influence becomes significant only within nonlinear overdensities where $\rho \gg \bar{\rho}$.

The model therefore interpolates smoothly between pressureless behavior at cosmological densities and pressure-supported equilibrium in collapsed halos. For the calibrated parameter range producing kiloparsec-scale cores, the predicted Jeans length remains sufficiently small to preserve large-scale structure formation within current observational bounds.

The computed matter power spectrum confirms that deviations from $\Lambda$CDM appear only at sufficiently large wavenumbers corresponding to small physical scales, while the large-scale spectrum remains essentially unchanged. In addition, the distribution of logarithmic rotation-curve slopes derived from SPARC galaxies supports the assumption of approximately constant velocity dispersion in halo interiors used in the effective fluid description.

These results indicate that the proposed framework preserves the successful large-scale predictions of $\Lambda$CDM while allowing controlled modifications to the inner structure of dark matter halos.

\section{Parameter Space and Astrophysical Constraints}

\subsection{Calibration from Dwarf Halo Cores}

The characteristic core scale for the $n=2$ polytropic configuration satisfies
\begin{equation}
a^2
=
\frac{3K}{4\pi G}
\rho_c^{-1/2}.
\end{equation}

Solving for $K$ in terms of the core radius $R_c \sim a$ gives
\begin{equation}
K
=
\frac{4\pi G}{3}
R_c^2
\rho_c^{1/2}.
\label{eq:K_calibrated}
\end{equation}

For representative dwarf spheroidal parameters,
\begin{equation}
R_c \sim 0.5\text{--}1 \,\mathrm{kpc},
\qquad
\rho_c \sim 0.05\text{--}0.2 \, M_\odot \,\mathrm{pc}^{-3},
\end{equation}
one obtains
\begin{equation}
K \sim 0.05\text{--}0.3
\quad
(\mathrm{km\,s^{-1}})^2
(M_\odot\,\mathrm{kpc}^{-3})^{-1/2}.
\end{equation}

Thus kiloparsec-scale cores correspond to values of $K$ of order $10^{-1}$ in the adopted units. Smaller values of $K$ approach the collisionless cold dark matter limit, whereas larger values produce progressively larger core radii.

\begin{figure}
    \centering
    \includegraphics[width=0.5\linewidth]{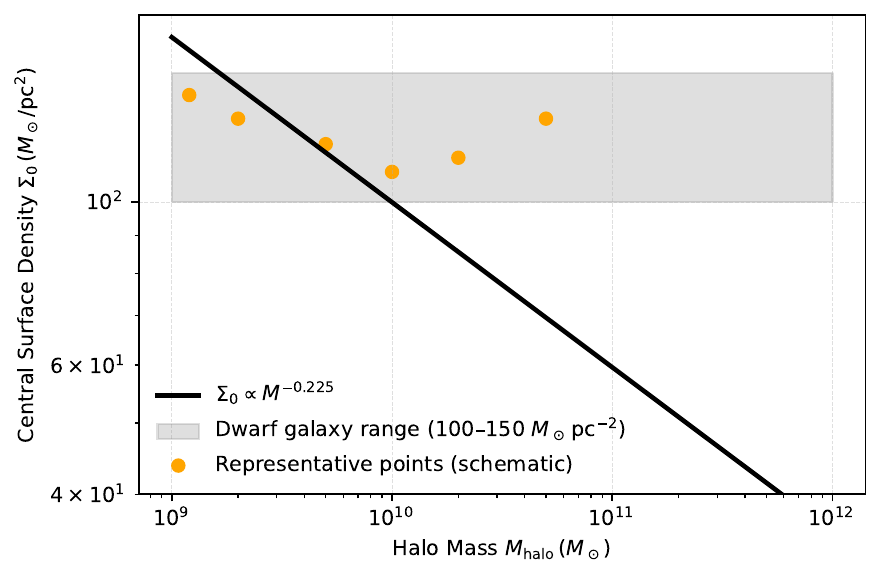}
    \caption{
Core radius $R_c$ as a function of the polytropic parameter $K$ for a fiducial dwarf central density $\rho_c \sim 0.1\,M_\odot\,\mathrm{pc}^{-3}$. 
The horizontal dashed lines indicate kiloparsec-scale cores, while the vertical dashed line marks the value of $K$ required to produce $R_c \approx 1\,\mathrm{kpc}$. 
The scaling $R_c \propto K^{1/2}$ follows directly from the Lane--Emden relation.
}
    \label{fig:placeholder}
\end{figure}

\subsection{Jeans Scale and Linear Structure Constraint}

The linear perturbation equation implies a comoving Jeans wavenumber
\begin{equation}
k_J^2
=
\frac{4\pi G \bar{\rho}}{c_s^2},
\qquad
c_s^2 = \gamma K \bar{\rho}^{\gamma - 1}.
\end{equation}

For $\gamma = 1.5$,
\begin{equation}
c_s^2 = 1.5 K \bar{\rho}^{1/2}.
\end{equation}

Using the present-day mean matter density
$\bar{\rho}_0 \simeq 38 \, M_\odot \,\mathrm{kpc}^{-3}$,
the Jeans wavenumber scales as
\begin{equation}
k_J \propto K^{-1/2}.
\end{equation}

For $K \sim 0.1$, corresponding to $R_c \sim 1\,\mathrm{kpc}$, one finds
\begin{equation}
k_J \approx 60\text{--}70 \, h\,\mathrm{Mpc}^{-1},
\end{equation}
which corresponds to a Jeans length
\begin{equation}
\lambda_J \sim 0.1\,\mathrm{Mpc}.
\end{equation}

This scale lies well below the wavenumbers probed by large-scale structure measurements, which typically constrain $k \lesssim 1\,h\,\mathrm{Mpc}^{-1}$. Consequently, linear growth and galaxy clustering at observable scales remain unaffected within the calibrated parameter range.

\subsection{Upper and Lower Bounds}

Two limiting regimes can be identified:

\begin{itemize}
\item \textbf{Small $K$:} For $K \ll 0.01$, the core radius falls below $\sim 0.1\,\mathrm{kpc}$ and the model approaches the pressureless cold dark matter limit.

\item \textbf{Large $K$:} For $K \gtrsim 1$, the Jeans scale moves toward nonlinear halo scales, potentially suppressing small-scale structure more strongly. Such values do not significantly modify currently constrained large-scale structure observables, although they may affect halo abundance at subgalactic masses.
\end{itemize}

Combining halo calibration with cosmological consistency yields an indicative viable range
\begin{equation}
0.01 \lesssim K \lesssim 1,
\end{equation}
within which kiloparsec-scale cores are produced while linear cosmology remains effectively unchanged.

Using representative dwarf-halo normalization
$\rho_{c,0} \simeq 0.1\,M_\odot\,\mathrm{pc}^{-3}$ and
$R_{c,0} \simeq 1\,\mathrm{kpc}$ at
$M_0 = 10^{10} M_\odot$, the model predicts

\begin{equation}
\Sigma_0(M)
\simeq
100
\left(
\frac{M}{10^{10} M_\odot}
\right)^{-0.225}
M_\odot\,\mathrm{pc}^{-2}.
\end{equation}

For dwarf-scale systems this yields
$\Sigma_0 \sim 100$--$150\,M_\odot\,\mathrm{pc}^{-2}$,
which is comparable to empirical central surface density estimates within current observational scatter.

\subsection{Summary of Constraints}

Within the above parameter range:

\begin{itemize}
\item Core radii of order $0.5$--$1\,\mathrm{kpc}$ arise naturally in dwarf-mass halos,
\item The present-day Jeans scale remains at $k_J \gg 1\,h\,\mathrm{Mpc}^{-1}$,
\item Large-scale clustering and weak lensing at linear scales remain unchanged,
\item Deviations from standard $\Lambda$CDM predictions are confined to nonlinear halo interiors.
\end{itemize}

These estimates demonstrate that a non-empty and observationally viable parameter window exists without requiring fine tuning.

\section{Weak Lensing Signatures}

\subsection{Projected Surface Density}

The observable quantity in weak gravitational lensing \cite{LensingReview}is the projected surface mass density,
\begin{equation}
\Sigma(R)
=
2 \int_R^{R_{\rm halo}}
\frac{\rho(r)\, r \, dr}{\sqrt{r^2 - R^2}},
\label{eq:Sigma_def}
\end{equation}
where $R$ denotes the projected radial coordinate in the lens plane. For the $n=2$ polytropic solution derived in Sec.~III, the density remains finite at $r=0$ and vanishes at $R_{\rm halo} = a \xi_1$.

Near the center, where
\[
\rho(r) \simeq \rho_c \left(1 - \frac{r^2}{6a^2}\right)^2,
\]
the integral in Eq.~(\ref{eq:Sigma_def}) yields
\begin{equation}
\Sigma(0)
=
2 \int_0^{R_{\rm halo}} \rho(r) \, dr
\sim
2 \rho_c a \, \mathcal{O}(1),
\label{eq:Sigma_center}
\end{equation}
which is finite and determined by the central density and the core scale $a$. The absence of a divergent inner slope ensures that the projected surface density remains bounded at $R=0$, in contrast to cuspy profiles whose projected density grows logarithmically toward the center.

\subsection*{Numerical Comparison with NFW}

To quantify the magnitude of the core-induced modification, we construct a smooth hybrid density profile in which the inner $n=2$ polytropic solution is continuously matched onto an outer Navarro--Frenk--White (NFW) halo. The transition is implemented over a narrow radial interval centered at $r_t \sim \mathcal{O}(R_c)$ using a smooth weighting function, ensuring continuity of the density profile while preserving the asymptotic NFW behavior at large radii.

The virial mass $M_{\rm vir}$ and concentration parameter $c$ are held fixed and identical for both configurations. The central density $\rho_c$ of the polytropic core is determined by requiring exact equality of the total enclosed mass, so that the hybrid and NFW halos are strictly mass-matched.

\begin{figure}
    \centering
    \includegraphics[width=0.5\linewidth]{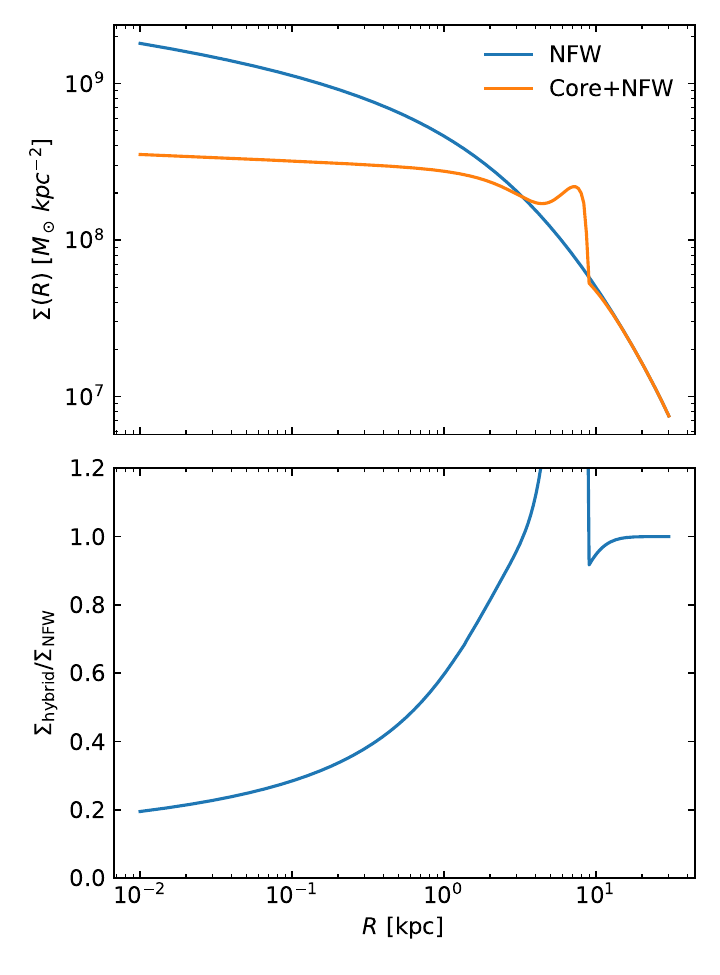}
\caption{
Top panel: Projected surface density $\Sigma(R)$ for a smooth core+NFW hybrid halo (solid orange curve) compared to a standard Navarro--Frenk--White (NFW) halo (blue curve) of identical virial mass $M_{\rm vir}$ and concentration $c$. The inner region is described by the $n=2$ polytropic solution ($\gamma=3/2$), smoothly matched onto the NFW profile at $r_t \sim \mathcal{O}(R_c)$.
Bottom panel: Ratio $\Sigma_{\rm hybrid}/\Sigma_{\rm NFW}$. The hybrid configuration exhibits suppression of the central projected density within $R \lesssim \text{few}\,R_c$, while converging to the NFW profile at larger radii. Deviations are confined to the nonlinear inner region, and the total halo mass is exactly matched between the two models.
}
    \label{fig:sigma_comparison}
\end{figure}

Figure~\ref{fig:sigma_comparison} shows the projected surface density $\Sigma(R)$ for the hybrid profile compared to a standard NFW halo with identical $M_{\rm vir}$ and $c$. The lower panel displays the ratio $\Sigma_{\rm hybrid}/\Sigma_{\rm NFW}$.

The hybrid configuration exhibits a finite central surface density and a monotonic suppression relative to the NFW profile for $R \lesssim \text{few}\,R_c$, while converging smoothly to the standard NFW behavior at larger radii. The deviation remains confined to the nonlinear inner region and leaves the outer mass distribution unchanged. This numerical comparison confirms the analytic expectation that the polytropic core modifies only the central halo structure while preserving large-scale consistency with $\Lambda$CDM.

\subsection{Convergence Profile}

The dimensionless lensing convergence is defined as
\begin{equation}
\kappa(R)
=
\frac{\Sigma(R)}{\Sigma_{\rm crit}},
\end{equation}
where the critical surface density is
\begin{equation}
\Sigma_{\rm crit}
=
\frac{c^2}{4\pi G}
\frac{D_s}{D_d D_{ds}},
\end{equation}
with $D_d$, $D_s$, and $D_{ds}$ denoting the angular-diameter distances to the deflector, the source, and between them.

Because $\Sigma(0)$ is finite, the convergence also satisfies
\begin{equation}
\kappa(0) < \infty.
\end{equation}

Matching the halo mass between the polytropic configuration and an NFW halo fixes $\rho_c$ and $a$ relative to $\rho_s$ and $r_s$. For representative dwarf-scale parameters, the central projected density of the polytropic halo is moderately suppressed relative to that of an equal-mass NFW halo. The precise ratio depends on the halo mass and concentration, but remains controlled by a single parameter $K$. Deviations are confined primarily to radii $R \lesssim R_c$, while the profiles converge toward one another at larger radii.

\subsection{Convergence Power Spectrum}

Weak-lensing surveys probe not only individual halo profiles but also the convergence power spectrum\cite{HaloModel} \cite{KilbingerReview},
\begin{equation}
P_\kappa(\ell)
=
\int_0^{\chi_H}
\frac{W^2(\chi)}{\chi^2}
P_m\!\left(k=\frac{\ell}{\chi},z(\chi)\right)
d\chi,
\label{eq:Pkappa}
\end{equation}
where $P_m(k)$ is the matter power spectrum and $W(\chi)$ is the lensing efficiency kernel.

Since linear growth remains unchanged for $k \ll k_J$, modifications arise primarily from the nonlinear one-halo term. The presence of a finite-density core introduces a characteristic Fourier scale
\[
k_c \sim \frac{1}{R_c}.
\]

For $kR_c \ll 1$, the leading correction to the one-halo contribution scales approximately as
\begin{equation}
\frac{\Delta P_\kappa}{P_\kappa}
\sim
- \frac{\ell^2 R_c^2}{\chi^2}.
\label{eq:lensing_scaling}
\end{equation}

Embedding the core scaling within the empirical concentration--mass relation $c(M) \propto M^{-\beta}$ yields
\begin{equation}
R_c(M) \propto M^{\frac{3\beta}{2}(2-\gamma)},
\end{equation}
and therefore
\begin{equation}
\frac{\Delta P_\kappa}{P_\kappa}
\propto
- \ell^2 M^{3\beta(2-\gamma)}.
\end{equation}

For $\beta \simeq 0.1$ and $\gamma \simeq 1.5$, the mass dependence is weak and deviations remain confined to sufficiently high multipoles.

\subsection{Constraint from Current Weak-Lensing Scales}

Current wide-field surveys measure the convergence power up to multipoles $\ell \sim 3000$ \cite{DES,HSC}. Requiring that the fractional suppression satisfies
\[
|\Delta P_\kappa / P_\kappa| < 0.1
\]
at this scale implies
\begin{equation}
R_c \lesssim 150 \,\mathrm{kpc},
\end{equation}
for characteristic lensing distances $\chi \sim 1500\,\mathrm{Mpc}$.

The kiloparsec-scale cores predicted in the parameter range considered here ($R_c \sim 1$--$5\,\mathrm{kpc}$) lie far below this bound. Consequently, current weak-lensing surveys do not impose significant constraints on $\gamma$ within the dynamically stable regime $\gamma > 4/3$. The predicted suppression therefore remains confined to scales beyond the reach of present measurements.

\subsection{Observational Outlook}

The model does not predict large-scale deviations from $\Lambda$CDM. Instead, it produces controlled modifications to the inner halo structure and correspondingly modest changes to the high-$\ell$ convergence power spectrum. These effects can be tested, in principle, with future high-resolution lensing surveys, although the present data do not yet probe the relevant multipole range with sufficient precision \cite{RomanMission}.

\section{Conclusion}

We have examined a minimal thermodynamic extension of cold dark matter characterized by a polytropic equation of state $P = K \rho^{\gamma}$ with $\gamma = 1.5$. The framework preserves the standard $\Lambda$CDM behavior at the background and linear perturbation levels while modifying the halo equilibrium in the nonlinear regime.

The resulting solutions exhibit finite-density, mass-dependent cores and a mild suppression of central lensing convergence relative to cuspy profiles of equal mass. The deviations are controlled by a single parameter and vanish smoothly in the pressureless limit $K \rightarrow 0$.

Although a full assessment ultimately requires numerical simulations and detailed comparison with observational data, the analytic results presented here demonstrate that a small effective pressure can modify inner halo structure without disrupting large-scale cosmology. Future high-resolution weak-lensing surveys may provide a promising avenue for testing these nonlinear signatures.
\appendix

\section{Lane--Emden Scaling and Perturbation Dynamics}

\subsection{Derivation of the Lane--Emden Equation for $n=2$}

We consider a static, spherically symmetric configuration satisfying hydrostatic equilibrium,
\begin{equation}
\frac{dP}{dr}
=
- \rho(r) \frac{G M(r)}{r^2},
\label{eq:hydrostatic_appendix}
\end{equation}
with enclosed mass
\begin{equation}
M(r)
=
4\pi \int_0^r \rho(r') r'^2 dr'.
\label{eq:mass_def_appendix}
\end{equation}

Adopting the polytropic equation of state
\begin{equation}
P = K \rho^{\gamma},
\qquad
\gamma = 1 + \frac{1}{n},
\end{equation}
and defining
\begin{equation}
\rho(r) = \rho_c \theta^n(\xi),
\qquad
r = a \xi,
\end{equation}
with scaling parameter
\begin{equation}
a^2
=
\frac{(n+1)K}{4\pi G}
\rho_c^{\frac{1}{n}-1},
\label{eq:a_scaling_appendix}
\end{equation}
one obtains, after substitution into Eq.~(\ref{eq:hydrostatic_appendix}), the dimensionless Lane--Emden equation,
\begin{equation}
\frac{1}{\xi^2}
\frac{d}{d\xi}
\left(
\xi^2 \frac{d\theta}{d\xi}
\right)
=
- \theta^n.
\label{eq:lane_emden_appendix}
\end{equation}

For $n=2$, the equation becomes
\begin{equation}
\frac{1}{\xi^2}
\frac{d}{d\xi}
\left(
\xi^2 \frac{d\theta}{d\xi}
\right)
=
- \theta^2,
\end{equation}
with boundary conditions
\begin{equation}
\theta(0)=1,
\qquad
\theta'(0)=0.
\end{equation}

The first zero of $\theta(\xi)$ occurs at
\begin{equation}
\xi_1 \approx 4.35,
\end{equation}
which defines the finite halo radius
$R_{\rm halo} = a \xi_1$.

\subsection{Mass--Core Scaling}

The total halo mass can be written as
\begin{equation}
M
=
4\pi a^3 \rho_c
\left[
- \xi^2 \theta'(\xi)
\right]_{\xi=\xi_1}.
\end{equation}

For $n=2$, the dimensionless coefficient
\begin{equation}
\omega_2
=
- \xi_1^2 \theta'(\xi_1)
\end{equation}
is a numerical constant. Hence,
\begin{equation}
M \propto a^3 \rho_c.
\end{equation}

Using Eq.~(\ref{eq:a_scaling_appendix}) with $n=2$,
\begin{equation}
a \propto K^{1/2} \rho_c^{-1/4},
\end{equation}
one obtains
\begin{equation}
M \propto K^{3/2} \rho_c^{1/4}.
\end{equation}

Eliminating $\rho_c$ yields
\begin{equation}
R_c \sim a \propto M^{-1/5},
\end{equation}
demonstrating the weak mass dependence of the core scale.

Equation~(A15) applies to isolated polytropic configurations. When embedded within the empirical concentration--mass relation of cosmological $\Lambda$CDM halos, this scaling is modified as discussed in Sec.~III.D.

\subsection{Linear Perturbation Equation}

In a homogeneous and isotropic background, the continuity and Euler equations for a barotropic fluid yield the standard density contrast evolution equation in Fourier space,
\begin{equation}
\ddot{\delta}
+
2H\dot{\delta}
=
4\pi G \bar{\rho} \delta
-
\frac{c_s^2 k^2}{a^2} \delta,
\label{eq:delta_appendix}
\end{equation}
where the adiabatic sound speed is
\begin{equation}
c_s^2
=
\frac{dP}{d\rho}
=
\gamma K \rho^{\gamma - 1}.
\end{equation}

For $\gamma=1.5$,
\begin{equation}
c_s^2
=
1.5 K \rho^{1/2}.
\end{equation}

The corresponding Jeans wavenumber satisfies
\begin{equation}
k_J^2
=
\frac{4\pi G \bar{\rho} a^2}{c_s^2},
\end{equation}
which scales as
\begin{equation}
k_J \propto K^{-1/2}.
\end{equation}

\subsection{Asymptotic Behavior}

Near the origin, the Lane--Emden solution admits the series expansion
\begin{equation}
\theta(\xi)
=
1 - \frac{\xi^2}{6}
+ \mathcal{O}(\xi^4),
\end{equation}
which implies
\begin{equation}
\rho(r)
=
\rho_c
\left(
1 - \frac{r^2}{6a^2}
\right)^2
+
\mathcal{O}(r^4).
\end{equation}

Thus,
\begin{equation}
\frac{d\ln\rho}{d\ln r}
\rightarrow 0
\qquad
(r \to 0),
\end{equation}
confirming the presence of a finite-density core.

At large radii approaching $\xi_1$, the density decreases rapidly toward zero,
\begin{equation}
\rho(r)
\propto
(\xi_1 - \xi)^2,
\end{equation}
leading to a sharp but finite halo boundary.

These asymptotic behaviors ensure that deviations from collisionless cuspy profiles remain confined to the inner halo while preserving a finite total mass.

\end{document}